\documentclass[prb,showpacs,superscriptaddress,floatfix,longbibliography,nofootinbib]{revtex4-2}
\usepackage{graphicx,amsfonts,amssymb,amsmath,xspace,dsfont}
\usepackage[colorlinks=true,citecolor=green,linkcolor=red,urlcolor=blue]{hyperref}
\usepackage{subfigure}
\usepackage{blindtext}
\usepackage{color}
\definecolor{g}{rgb}{.1,0.4,.1} 
\definecolor{b}{rgb}{0,0.2,1}
\definecolor{rouge}{rgb}{0.82,0.,0.}
\definecolor{vert}{rgb}{0.,0.82,0.}
\definecolor{orange}{rgb}{1,0.5,0.}
\definecolor{bleu}{rgb}{0.,0.,0.82}
\definecolor{m}{rgb}{0.82,0.,0.82}
\definecolor{vert2}{rgb}{0.,0.5,0.}
\definecolor{rougeclair}{rgb}{1.0,0.7,0.7}

\newcommand{\beq}{\begin{equation}}
\newcommand{\be}{\begin{equation}}
\newcommand{\beqn}{\begin{eqnarray}}
\newcommand{\eeq}{\end{equation}}
\newcommand{\ee}{\end{equation}}
\newcommand{\eeqn}{\end{eqnarray}}

\begin{document}
\title{Some attempts toward 3-dimensional Phyllotaxy}
\author{ R\'emy Mosseri}
\affiliation{Laboratoire de Physique Th\'eorique de la Mati\`ere Condens\'ee, Sorbonne Universit\'e, CNRS UMR 7600, F-75005 Paris, France}
\author{Jean-Fran\c cois Sadoc}
\affiliation{Laboratoire de Physique des Solides (CNRS-UMR 8502), B{\^a}t. 510, Universit{\'e} Paris-sud (Paris-Saclay), F 91405 Orsay Cedex, France}

\begin{abstract}
This paper investigates several distinct attempts to generalize in higher dimension the standard 2-dimensional phyllotaxy set construction. We first recall known contructions for these sets on $2D$ manifolds of constant curvature (the Euclidean plane $\mathbb{R}^2$, the sphere $\mathbb{S}^2$ and the hyperbolic plane $\mathbb{H}^2$). We then propose a first attempt to get a $3D$ phyllotactic set by piling up suitably shifted Euclidean $2D$ phyllotactic sets. A different, radially triggered, solution is then analyzed. An interesting phyllotactic set on the hypersphere $\mathbb{S}^3$ is then generated using a Hopf fibration approach. Finally,a simple 4-dimensional example is presented, generated as a simple product of two 2-dimensional planar sets. A $3D$ phyllotaxy candidate is then derived by applying a "Cut and Project" algorithm.
\end{abstract}
\maketitle

\section{Introduction}

Phyllotaxy, originally devoted to the study of leaves and florets arrangements in plants, leads to  interesting $2D$ spiral sets of points, whose geometry depends crucially on the golden mean irrational number (see Ref.~\cite{jean1983,jean1992,adlerbarabejean1997} for general overviews). Although the generating algorithm is rather simple (along a so-called ``Fermat spiral"), the properties of the obtained point set are quite subtle, with neighbouring points arranged along secondary spirals (the parastiches) interrupted by dislocations \cite{rivier1986}\cite{sadocriviercharvolin2012}
 
Seen as ``regular" arrangements of points in space, phyllotactic sets  are one solution one among many others (see Ref~\cite{mackay1986}), ranging from crystalline structures to quasicrystalline ones~\cite{schechtman1984} or amorphous and frustrated systems~\cite{zallen1983}\cite{sadocmosseri1999}.


Living systems also display regularity (and most often chirality), but not generically of periodic type. As examples, we find at a macromolecular level the DNA double helix , liquid crystalline-like structures in all sorts of biological membranes , and also, at a more macroscopic level, the above-mentioned plant phyllotaxy question. Notice that phyllotaxy has also been invoked in the macromolecular level context~\cite{charvolinsadoc2011}.

Our aim here is to consider deterministic arrangements that arise from simple phyllotactic sequences, leading to disk packings in $2D$ and eventually sphere packings in $3D$. On constant curvature 2-dimensional manifolds, these arrangements are generated by discrete dynamical systems on top of generative spirals, which will be recalled here. Our main purpose here is to try a generalization to three dimensions.

Notice that there is an important scientific literature devoted to the relation between these simple arrangements and real plant phyllotaxy; this is left to experts and will not be discussed here. There have been in addition some attempts to describe how simple physical rules can select these arrangements~\cite{rivierkoch1991,levitov1991,douady1992,douady1996,leelevitov1998}.

In a first step, we shall recall the  known phyllotaxy algorithms for generating sets in 2-dimensional manifold of constant curvaure (plane $\mathbb{R}^2$, sphere $\mathbb{S}^2$ and hyperbolic plane $\mathbb{H}^2$).

We then propose some ways to address the 3-dimensional case, with examples in $\mathbb{R}^3$ and $\mathbb{S}^3$. This part, although new, is still exploratory and will require further analysis.
A simple 4-dimensional case will also be presented, and we shall finally, in Appendix, display some curious and unexpected 3-dimensional figures that were accidentally generated while playing with some parameters.

\section{2-dimensional phyllotaxy on constant curvature homogeneous spaces}

2-dimensional Euclidean packing problems have been widely studied, either for crystalline, quasiperiodic or disordered arrangements. Phyllotactic arrangements are quite special in that they allow for simple deterministic sets which present some nice homogeneity properties. Quite generally, sites in a phyllotaxy point set belong to a generating spiral, along which they are characterized by  integers. The radial part of the coordinate depends on the $2D$ space curvature while the azimuthal angle is an irrational rotation (generically associated with the golden mean).

\subsection{The Euclidean planar case}
\label{subsec:euclideanr2}

The standard Euclidean $2D$ phyllotaxy consists of the generation  of an infinite discrete sequence of points, called $P(\mathbb{R}^2,\lambda, s)$, given here in terms of   complex coordinates $q= x + i y$ :

\begin{eqnarray}
q(s) =a \sqrt{s} \exp  (2 i \pi s \lambda), \nonumber \\
 \text{with} \; \lambda= 1/\tau \; \text{and} \; \tau=\frac{1+\sqrt{5}}{2}
\label{eq:phylloR2}
\end{eqnarray}

\begin{figure}
\begin{center}
\includegraphics[width=0.495\textwidth]{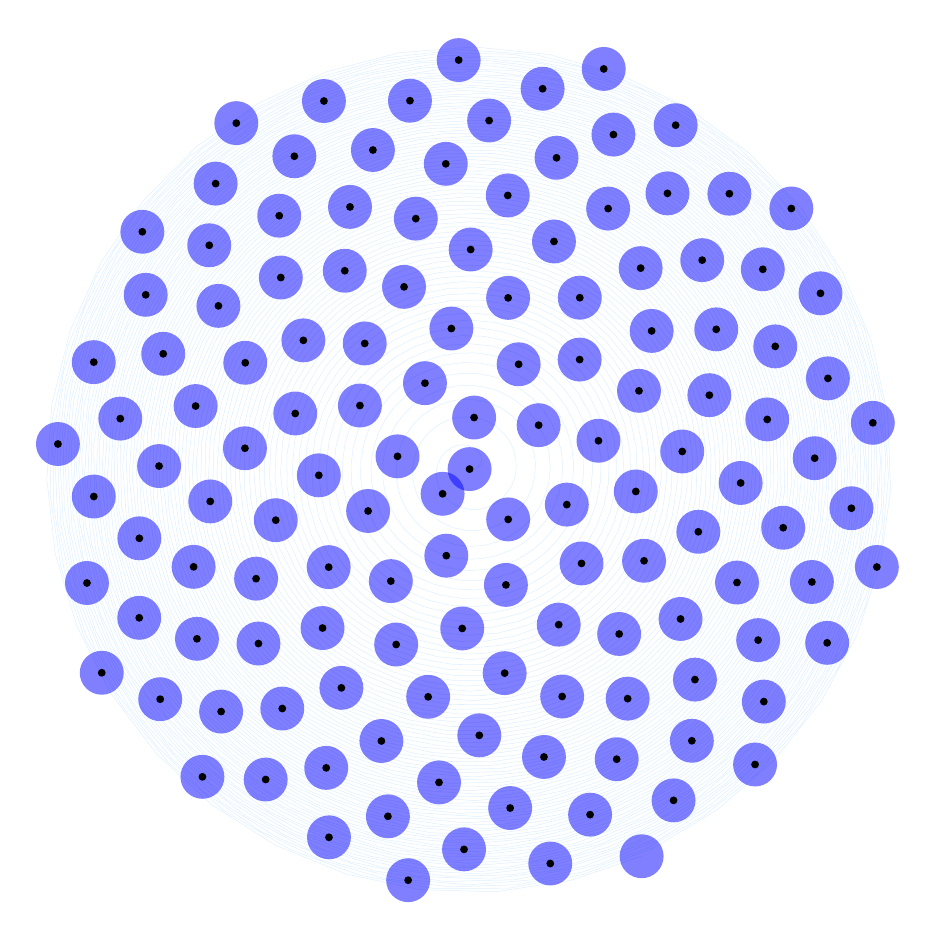}
\end{center}
\caption{Phyllotaxy on $\mathbb{R}^2$, with disks centred on the generative spiral.}
\label{fig:phylloR2}
\end{figure}

$\lambda$ characterizes the irrational rotation, which is best taken as the inverse of the golden mean to maximize the homogeneity of the generated sets (and will be omitted). The successive sites belong to a so-called Fermat spiral.
$a$ is a simple global scaling parameter, and $s$ are integers labelling the generated sites; we will sometime find usefull to shift $s$ by a real number $r<1$, leading to a translation along the generating spiral, and a set $P(\mathbb{R}^2,s+r)$ (see Fig.\ref{fig:phylloR2}). The $s^{1/2} $ factor in front ensures that each site adds a constant area to the cluster.  More interestingly, each site numbered $s$ (having on average 6 neighbours) has its neighbouring sites labelled by $s \pm F_j$ with $F_j$ a Fibonacci number. These neighbours belong to secondary spirals, called ``parastiches". The three sets of Fibonacci numbers are constant inside an annulus-like region. Shift to new parastiches given by new Fibonacci numbers occurs in ``defect-like" regions which have been characterized as dislocations~\cite{sadocriviercharvolin2012}

%

Dressing the $P(\mathbb{R}^2,s)$ sites with disks leads to a homogeneous planar disk packing , quite less dense than the triangular lattice disk packing. Note however that, contrary to the lattice case, owing to the irrationality of $\lambda$, all $P(\mathbb{R}^2,s)$ sites are visible from the origin.

\subsection{Phyllotaxy on the sphere $\mathbb{S}^2$}
\label{subsec:phyllos22}

It is quite easy to generalize planar phyllotaxy to a unit radius sphere $\mathbb{S}^2$ (see\cite{gonzalez2010}\cite{sadoccharvolinrivier2013}). But since the sphere is compact, one should specify in addition the total number of sites, denoted $n$, that is aimed to generate the homogeneous sphere covering.  Coordinates on a unit radius $\mathbb{S}^2$ are given in term of the complex number $q=x+ i y$ and the third coordinate $z$ :

\begin{eqnarray}
q(s) &=& \frac{2}{n}\sqrt{s(n-s)} \,  \exp  (2 i \pi s \lambda) \nonumber \\
z(s)&=& (n-2s)/n 
\label{eq:phylloS2}
\end{eqnarray}

\begin{figure}
\begin{center}
\includegraphics[width=0.4\textwidth]{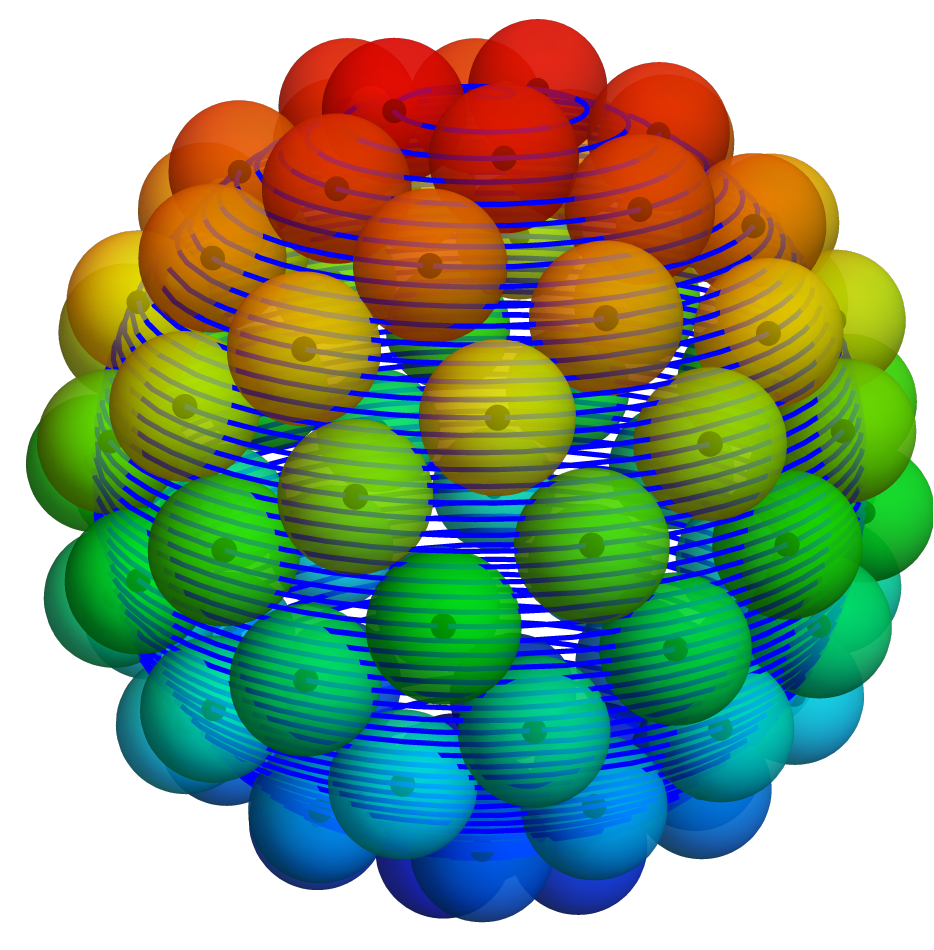}
\end{center}
\caption{Phyllotaxy $P(\mathbb{S}^2,s,F_{11})$ on the sphere $\mathbb{S}^2$, with $F_{11}=89$ sites represented by black small spheres, located on the underlying generative spiral, and surrounded by larger spheres whose colour goes from red (north pole) to blue (south pole) as $s$ increases.}
\label{fig:phylloS2}
\end{figure}
%

We choose here to generate sets with $n=F_j$ sites, leading to the sets denoted $P(\mathbb{S}^2,s,F_j)$ (see Fig.~\ref{fig:phylloS2}).
The generating spiral starts at the north pole for  $s=0$ and ends at the south pole for $s=n$. One can again define parastiches as secondary spirals connecting close neighbouring sites on the sphere.

Let us note a difference with the Euclidean case, where the $a$ factor is a global scaling factor (all generated sets are equivalent under scaling by $a$). The   $\mathbb{S}^2$ curved space has an intrinsic length scale (say the radius of curvature), and different phyllotaxy sets are generated according to this radius. Here this difference is labelled by the number of sites in the set.

\begin{figure}[h]
\begin{center}
\includegraphics[width=0.4\textwidth]{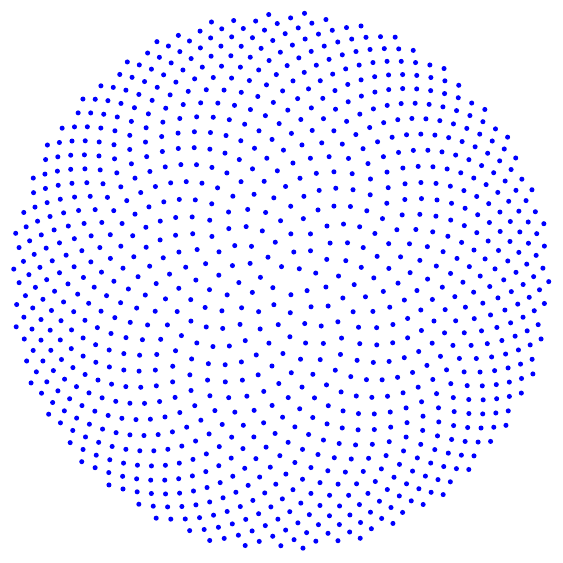}
\end{center}
\caption{Phyllotaxy on the hyperbolic plane $\mathbb{H}^2$ in the Poincar\'e disk representation. This set contains 1000 sites with a factor $a=1/50$. The portion of the Poincar/'e unit disk represented here is a small disk (of radius much smaller than unity). }
\label{fig:phylloH2}
\end{figure}
%

\subsection{Phyllotaxy on the negatively curved hyperbolic plane $\mathbb{H}^2$}
\label{subsec:phylloh2}

We recall here the presentation given in Ref.\cite{sadoccharvolinrivier2013}. A generative spiral is constructed on $\mathbb{H}^2$, which differs from the Euclidean version by its radial part. When displayed in the Poincar\'e disk representation (see Fig~\ref{fig:phylloH2}), the coordinates of $P(\mathbb{H}^2,s,a)$ reads

\begin{eqnarray}
q(s,a) &=& \rho (s,a) \exp  (2 i \pi s \lambda), \nonumber \\
\rho (s,a) &=& \tanh(\frac{1}{2} \cosh^{-1}(\frac{a^2 s}{2}+1))
\label{phylloH2}
\end{eqnarray}

 In this representation, the full negatively curved manifold is sent into the interior of a unique radius disk in $\mathbb{R}^2$, and the geodesics are circle arcs orthogonal to the unit circle. The factor $a$  is different here from its Euclidean counterpart and is linked to the radius of curvature. The sets $P(\mathbb{H}^2,a,s)$ with distinct $a$ are different.


\section{Toward $3D$ phyllotaxy in $\mathbb{R}^3$}
\label{phyllor3}

\subsection{Vertical packing of $2D$ phyllotaxy sets}
\label{phylloverticalr3}

Dense sphere packing in $\mathbb{R}^3$ can be simply described as vertical packings of 2-dimensional triangular array of spheres, such that each new sheet is translated such its spheres are placed on top of the largest interstice (triangle centre) of the preceding sheet. Calling $A$ a reference triangular plane position, there are two possible densest locations $B$ and $C$ for the addition on top of $A$. Face centred cubic (fcc) packings correspond to the sequence $ABCABC ...$, while hexagonal cubic (hc) packings correspond to the sequence $ABABAB...$. Besides these simple periodic lattices, one can form any equally dense non periodic  arrangements called "polytypes".

\begin{figure}[h]
\begin{center}
\includegraphics[width=0.4\textwidth]{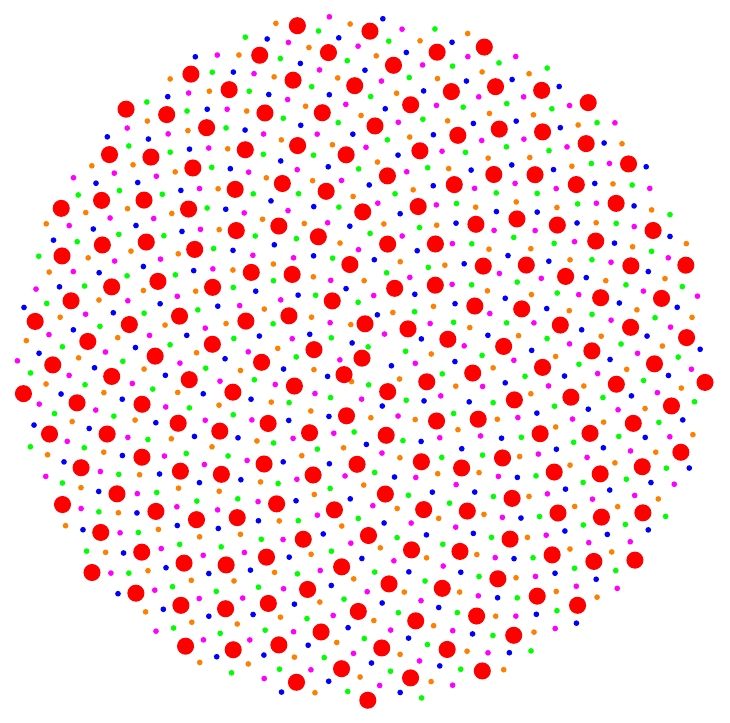}
\end{center}
\caption{Five sets $P(\mathbb{R}^2,s+j r)$ which nicely fill in the interstices created by each other. The set $j=0$ is represented by larger red disks, and $j=1,2,3,4$ are represented by smaller disks of respective colours green, blue, magenta and orange.}
\label{fig:phylloR2R3}
\end{figure}

As a first step toward 3-dimensional phyllotaxy, we are going to mimic the previous description and try to vertically pack suitably placed 2-dimensional phyllotaxy sets.
We start with a $P(\mathbb{R}^2,s)$ set, and check whether there exists a value $r$ such that $P(\mathbb{R}^2,s+r)$ sites fall nicely at the previous set interstices. One simple example (we do not claim it is unique) is given by $ r= \frac{3\lambda -1}{5\lambda}$. Repeating this shift, we generate eventually five sets $P(\mathbb{R}^2,s+j r)$ with $j=0...4$. These five sets are represented in Fig~\ref{fig:phylloR2R3}.

%

\begin{figure}
\begin{center}
\includegraphics[width=0.4\textwidth]{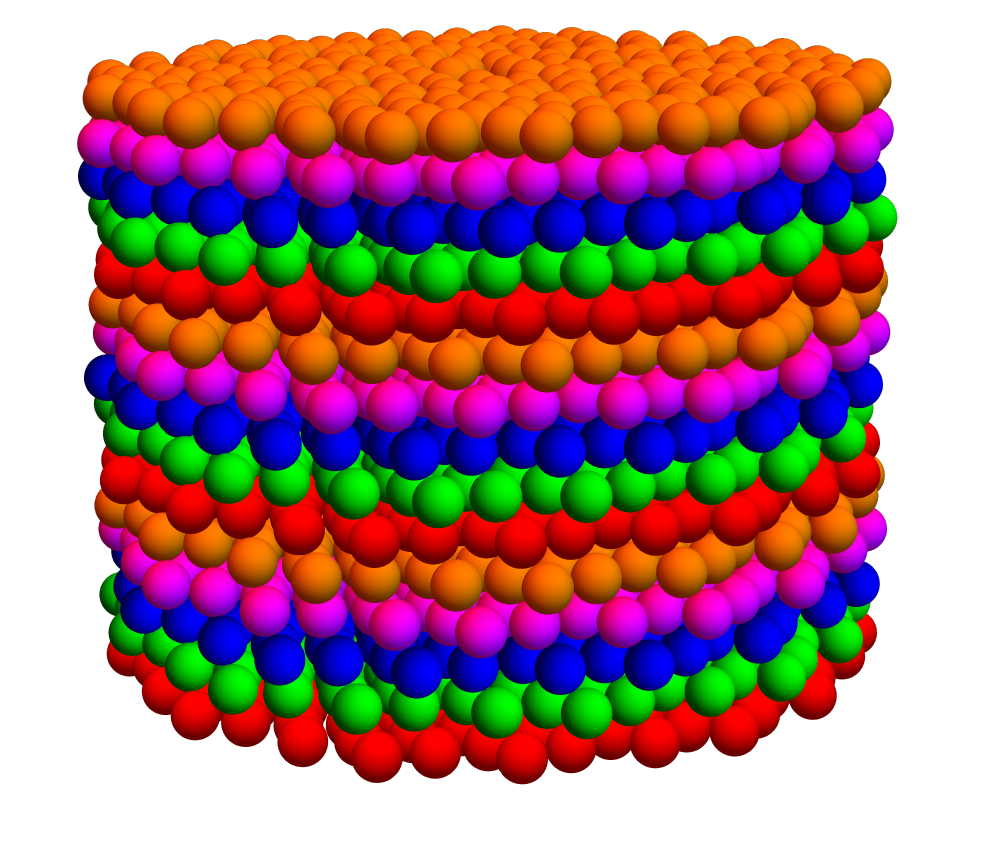}
\end{center}
\caption{A 3-dimensional phyllotaxy set $P(\mathbb{R}^3,s+ j r)$ We use the same colour choices as in Fig~\ref{fig:phylloR2R3}, with repeated colours corresponding to almost (but not exactly)identical sets vertically shifted by $5h$.}
\label{fig:phylloR3}
\end{figure}

The next set ($j=5$) almost covers sites of the $j=0$ set. Indeed, it is easily verified that the site $s$ for $j=5$ has exactly the same angular coordinate as the site $s+3$  for $j=0$, with a slightly different radial coordinate (at a ratio $\sqrt{(s+3)/3}$, a difference which vanishes as $s$ increases. The next step toward $3D$ phyllotaxy amounts to a vertical $j$-dependant shift and piling up these horizontal sets to get  the sphere arrangement, denoted $P(\mathbb{R}^3,s+j r)$, shown in Fig~\ref{fig:phylloR3}, with site coordinates (in the form $(q,z)$) :

\begin{eqnarray}
q(s,j) &=&a \sqrt{s+ \text{frac}(j r)} \exp  (2 i \pi ((s+j r) \lambda), \nonumber \\
z(j)&=&j h,
\label{eq:phylloR3}
\end{eqnarray}
with $h$ the interplane distance, and ``frac" is the fractional part.

%

An interesting next step will be to operate a (not yet done) Voronoi decomposition of $P(\mathbb{R}^3,s+ j r)$ sets to better characterize the local order and see how the $2D$ parastiches (seen as local neighbourhood markers) extend in $3D$. Due to the construction algorithm used here, we can conjecture the following : the $2D$ parastiches should extend to helicoidal $3D$ leaves, up to cylindrical grain boundaries made of dislocation lines leading to a next series of $3D$ parastiches of higher order.

\subsection{Tentative radial packing from  a $\mathbb{S}^{2}$ phyllotaxy set}
\label{subsec:radialr3}

Crystalline sphere packings can be decomposed into successive shells of increasing radius with symmetrical arrangements compatible with the crystal point group symmetry. A first na\"ive idea would be here to consider that $P(\mathbb{S}^2,\lambda, s)$ sets with increasing radius and  number of sites could realize an interesting  shelling in $3D$. However, this task would require rather precise matching conditions between the successive spherical shell radii, the number of sites on each shell, and possible intersphere phase shifts along the generative spirals. We did not follow this track, which would  nevertheless be kept as plausible.

\subsubsection{A deterministic approach}
\label{deterministic}

Another (quite distinct) possibility consists of starting from a unique generative template set $P(\mathbb{S}^2,\lambda, s)$  and displacing radially the sites to generate a $3D$ arrangement. Notice that the above $P(\mathbb{R}^2,\lambda, s)$ set could have been introduced in a similar way : start from an irrational rotation on the (generative set) unit circle $\mathbb{S}^1$, given by $q(s) = \exp  (2 i \pi s \lambda)$, and then move radially each point according to radial distance $\rho(s)= \sqrt{s}$. 

The simple relation between the rotational and radial parts allows for the deterministic construction of the phyllotaxy set in the $2D$ case. The situation is more complex if we try to apply a similar approach in $3D$, upon starting from a $P(\mathbb{S}^2,s,F_j)$ set and radially displace the site at a radial distance $\rho(s)= s^{1/3}$.  Even if the radial distance is fixed by the ordering, we now have two parameters, the polar (latitude) and azimuthal (longitude)  angles to specify the $s^{th}$ site. Another way to say that is that we are lacking an underlying generative spiral (which leads to a natural order among the sites) in $3D$.  This leads us to include a potential reordering of the sites, a permutation $s \rightarrow \, t(s)$, here done via a modulo application, in the form

\begin{eqnarray}
q(s) &=&  s^{1/3} \frac{2}{n}\sqrt{t(n-t)} \,  \exp  (2 i \pi t \lambda) \nonumber \\
z(s)&=&  s^{1/3} (n-2t)/n \quad \rm{with} \nonumber \\
t &=&  s\,m \, {\rm Mod} \, n \quad m \in \mathbb{N} \; {\rm and} \; (m,n) \; \rm{coprimes}
\label{eq:phylloS2R3}
\end{eqnarray}

\begin{figure}
\begin{center}
\includegraphics[width=1.0\textwidth]{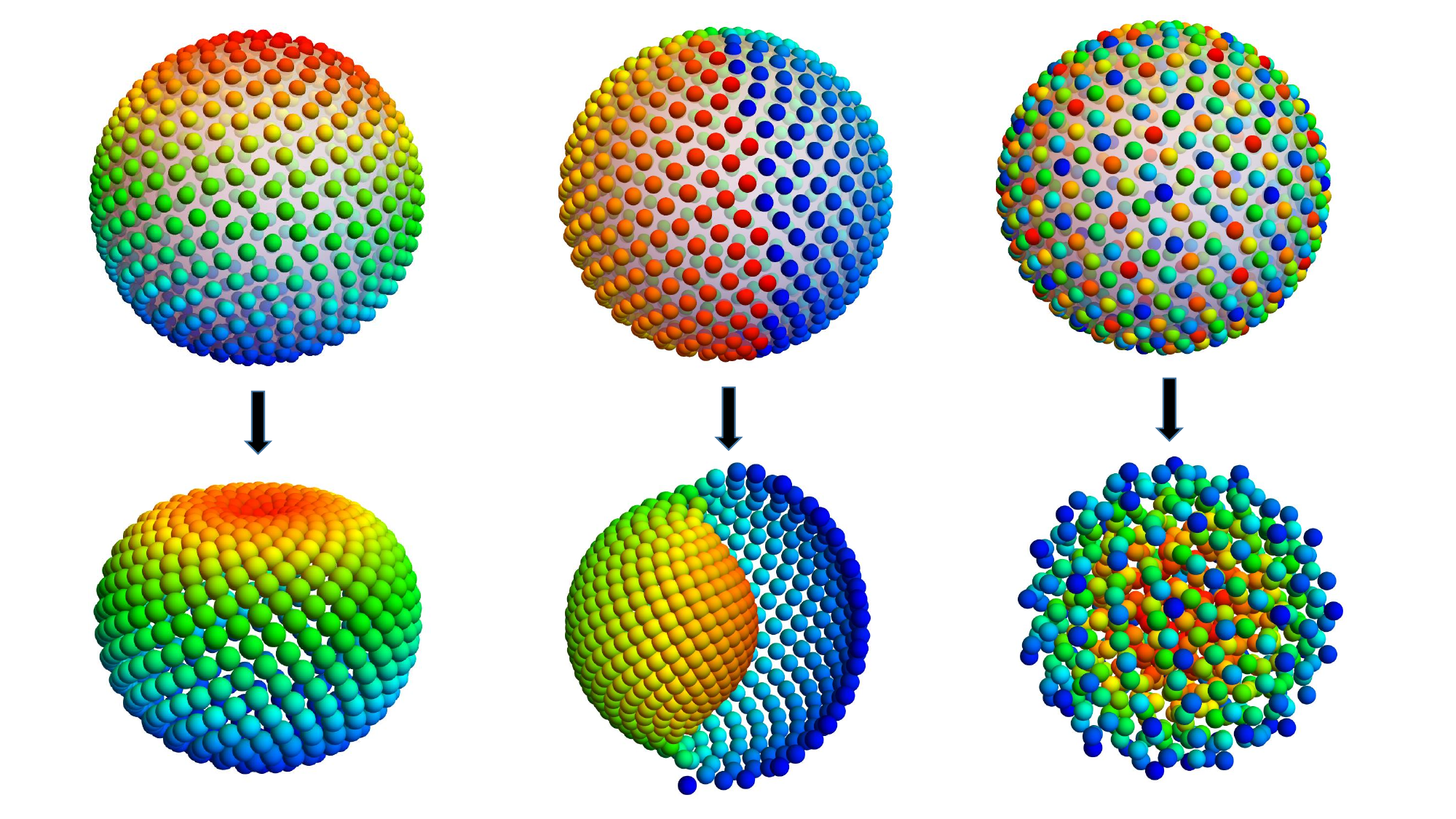}
\end{center}
\caption{ Deterministic sets derived from a template $P(\mathbb{S}^2,F_{15})$. Sites, reordered via the transformation $s \rightarrow t(s)$, are placed at a radial distance $ s^{1/3} $. Left : For $t(s)=s$ (latitude ordering on the generative set, see top left), one gets a deformed spherical topology near the north pole (lower left); Middle : We use the transformation $t(s)= s F_{14}$ Mod $F_{15}$. Sites in the template are now ordered according to their longitude (upper plot), leading to an unexpected shell-like shape (lower plot); Right : A  reordering leading to a more homogeneous $3D$ pattern (here we used $m=39$). The colour code goes from red to blue while increasing $s$.} 
\label{fig:deterministicS2R3}
\end{figure}

The condition for $(m,n)$ to be coprimes ensures that all sites in the generative template set are used to generate the $3D$ structure. The coordinates of $P(\mathbb{S}^2,\lambda, s)$ given in Eq.~\ref{eq:phylloS2} are such that  $s$ orders the sites according to their latitude in the template. As a result, the $3D$ structure obtained from  Eq.~\ref{eq:phylloS2R3} will retain a $2D$ substructure, as shown in Fig.\ref{fig:deterministicS2R3}-left. An interesting case corresponds to using the transformation 
$t(s)= s F_{j-1}$ Mod $F_j$. Here $t(s)$ orders sites according to their longitude, leading to a nice shell-like structure shown in Fig.\ref{fig:deterministicS2R3}-middle, displaying still a $2D$ substructure. In order to get rid of the latter, one needs to use a generic modulo transformation such that $t(s+1)$ selects a site located far enough from $t(s)$ in the generative template. Recalling that neighbouring sites  in the original ordering are along parastiches with sites number differing by a Fibonacci number, one should at least avoid choosing $m$ in the Fibonacci sequence. An example is shown in Fig.\ref{fig:deterministicS2R3}-right.

We have therefore shown that it is possible to propose an ordering in the template set such that the obtained deterministic structure extends more homogeneously in $3D$. We nevertheless do not pretend here to have found at this stage the ``best" reordering. 

The interested reader is invited to display the quite nice bestiary of structures obtained by varying $m$ in Eq.~\ref{eq:phylloS2R3}, which sometimes displays unexpected shapes. The $(m,n)$ coprime condition can also be relaxed to generate additional shapes. One can even decide to keep on generating the cluster  with $s>n$, leading to multi-shells arrangements.

Being intrigued by the nice shell structure displayed in Fig.~\ref{fig:deterministicS2R3}-middle, we further played with the parameters, leading to interesting other structures shown in the Appendix.

\subsubsection{A numerical approach}

As an alternative approach, we can try to numerically solve for $t(s)$ by iteratively generating a cluster $C(s)$ given by Eq.~\ref{eq:phylloS2R3}, but with the reordering $t(s)$ being now the result of an optimization process rather than following an automatic rule. The more natural one consists in  asking at each step that the new site $t(s+1)$, once brought at the radial distance $(s+1)^{1/3}$, fills the largest hole in the growing cluster. This notion  has already been invoked~\cite{douady1992} as an algorithmic rule to grow phyllotaxy sets in $2D$, and is usually implemented via  repulsive potentials centred on the already created sites. The difference here is that it is tried in $3D$, with new sites radially brought from a template set and placed at a precise distance. We therefore do not look for the precise local minimum of the pair potential, but rather expect to compare different possible site accretion. 
 We call $\bar C$ the complementary set (the sites in the template $P(\mathbb{S}^2,\lambda, s)$ which have not yet been incorporated in the growing cluster). At each step, all sites $t$ in $\bar C$ are successively placed at a radial distance $(s+1)^{1/3}$ and are compared with respect to a repulsive potential created by nearby sites in $C(s)$. The site $t_0$ which minimizes this interaction potential is then selected : $t(s+1)=t_0$.

\begin{figure}
\begin{center}
\includegraphics[width=1.0\textwidth]{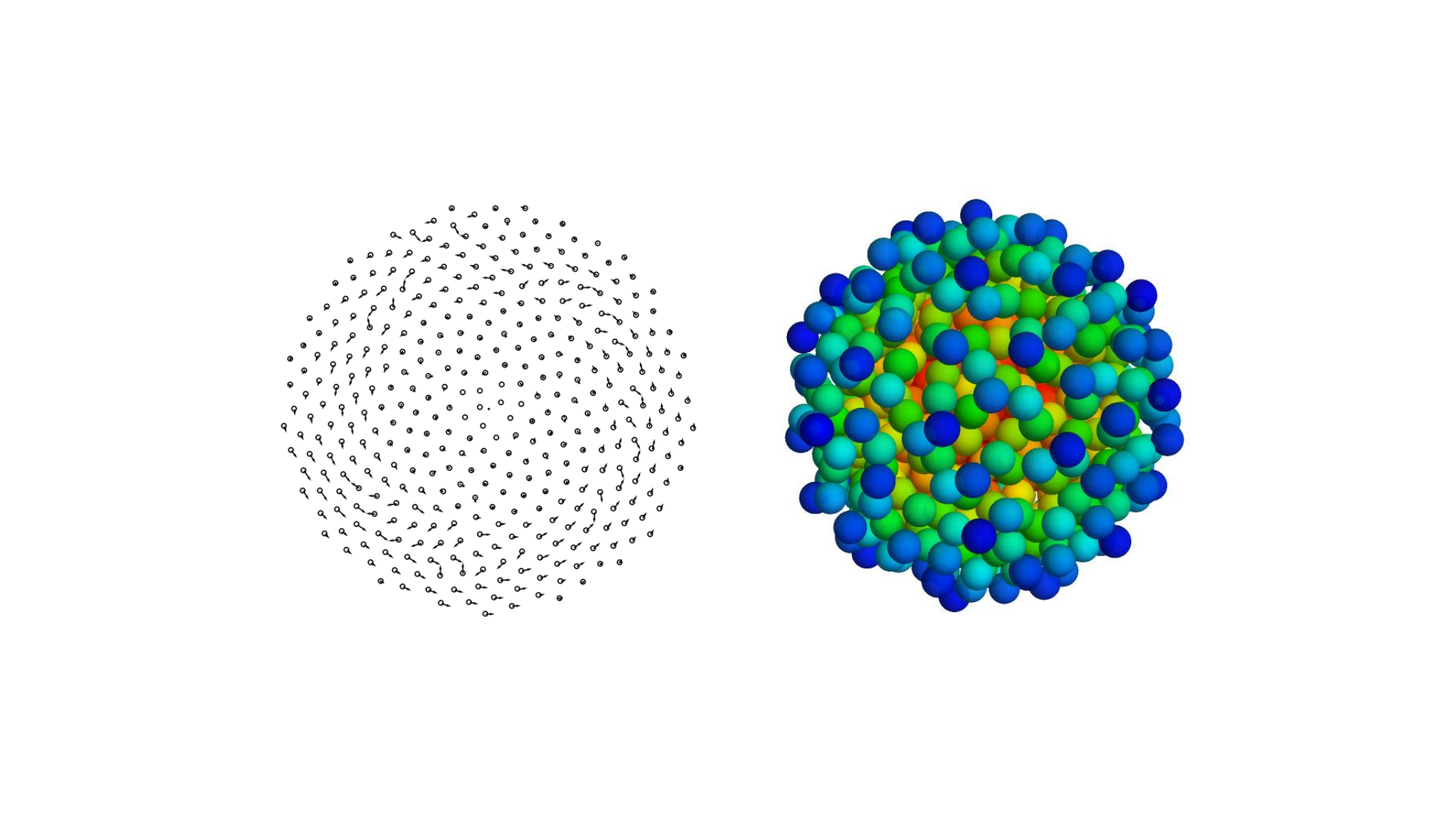}
\end{center}
\caption{ Numerically generated clusters : (i) Left , benchmark of the numerical approach in $2D$. A cluster with   $F_{15}=610$ sites has been generated. Arrows show the difference (for  $F_{14}=377$ sites to avoid boundary effects) between numerically derived positions and those given by Eq.~\ref{eq:phylloR2}; (ii) Right, numerically derived $3D$ cluster with $F_{15}=610$ sites generated with the same algorithm as in $2D$, with a  colour code from red to blue while increasing $s$.. 
}
\label{fig:phyllonumericR3}
\end{figure}
%

This approach can be benchmarked in $2D$. This test is presented in Fig.~\ref{fig:phyllonumericR3} on the left. The selected sites are not exactly located at the position given by Eq.~\ref{eq:phylloR2}, but appear not too far (the difference being shown as an arrow.  This is an invitation to apply this approach to the $3D$ case, leading an interesting 3$D$ phyllotaxy set, an example of which is shown in Fig.~\ref{fig:phyllonumericR3}-right.

\section{Phyllotaxy in $\mathbb{S}^3$}

The 3-sphere $\mathbb{S}^3$ is a curved $3D$ space, which can be embedded in 4D Euclidean space by the simple equation $x_1^2+x_2^2+x_3^2+x_4^2=1$. 
One should not forget that it is a (curved) 3-dimensional space (indeed,the preceding four coordinates are constrained by one equation). Let us have a closer look at $\mathbb{S}^{3}$ via the concept of fibred space. A fibred space $E$ is defined by a mapping $p$ from $E$ onto the so-called ``base" $B$, any point of a given fibre being mapped onto the same base point. A fibre is therefore the full pre-image of one base point under the mapping $p$. In a 3-dimensional fibred space with 1-dimensional fibres, the base is a 2-dimensional manifold. When the base manifold cannot be embedded in the full space, the fibration is said to be non trivial. This is the case for the Hopf fibration \cite{hopf1931}\cite{mosseri2012} of $\mathbb{S}^{3}$ by great circles $\mathbb{S}^{1}$ and base $\mathbb{S}^{2}$ for which $\mathbb{S}^{3}\neq \mathbb{S}^{2}\times \mathbb{S}^{1}$.

Let us use describe the unit radius $\mathbb{S}^{3}$ with complex coordinates
$q_1=x_{1}+\imath x_{2}$ and $q_2=x_{3}+\imath x_{4}$ satisfying $|q_1|^2+|q_2|^2=1$
. 

The Hopf map  may then be defined as the composition of a map $h_{1}$ from $\mathbb{S}^{3}$ to $\mathbb{R}^{2}$+$\{\infty\}$ followed by an inverse stereographic projection $h_{2}$ from $\mathbb{R}^{2}$ to the unit sphere $\mathbb{S}^{2}$, with coordinates $(x,y,z)$ reading:

\begin{eqnarray}
x &=& 2 \, \rm{Re}(q_1\bar{q_2})\nonumber \\
y &=& 2 \, \rm{Im}(q_1\bar{q_2}) \nonumber \\
z &=& |q_1|^2-|q_2|^2 
\label{eq:hopfmap}
\end{eqnarray}

where $\bar{q}$ denotes the complex conjugate of $q$.

The fibre is a great circle parametrized by one angle
 $\theta$: $(q_1,q_2)\longrightarrow (e^{{\bf i}\theta}q_1,e^{{\bf i}\theta}q_2)$;
clearly the above base points do not depend on $\theta$ and each one therefore characterizes
a fibre.

\begin{figure}
\includegraphics[width=1.0\textwidth]{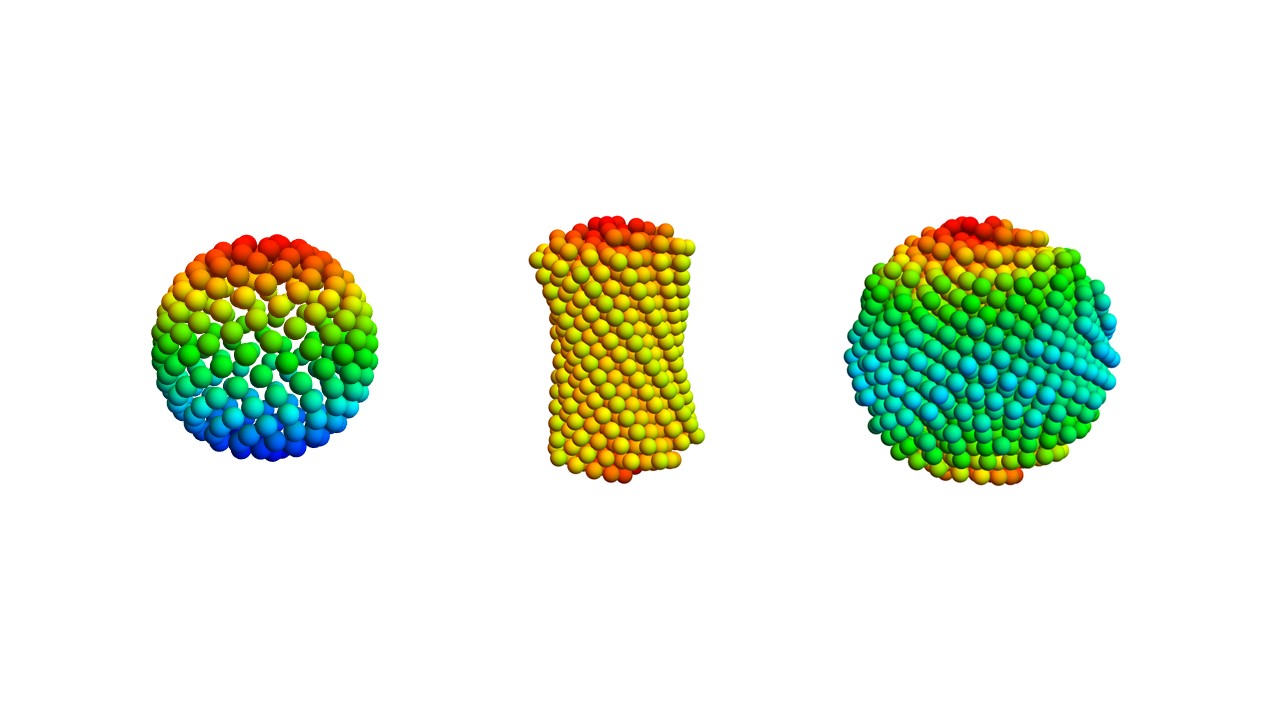}\hfill
\caption{Example of a $P(\mathbb{S}^3,F_{13})$ set. Left: Hopf map onto the $\mathbb{S}^2$ base of the $P(\mathbb{S}^3,F_{13})$ set (each point represents a fibre). Center :  Stereographic map (keeping only sites falling inside a given ball of radius 0.6 in $\mathbb{R}^3$) of a portion of the phyllotaxy set $P(\mathbb{S}^3,F_{13})$, with a selection of fibres near the north pole of the base. Sites belonging to the same fibre are equally coloured (according to the colour of the corresponding point on the base). Right : full stereographic projection of $P(\mathbb{S}^3,F_{13})$, keeping only sites falling inside a ball of radius 0.6 in $\mathbb{R}^3$.}
\label{fig:phylloS3R3}
\end{figure}
%

The idea is now the following : since any collection of points on the unit sphere $\mathbb{S}^2$ (viewed as the base of a Hopf fibration) defines a set of Hopf circular fibres on $\mathbb{S}^3$, we start with a phyllotactic set $P(\mathbb{S}^2,F_j)$. This leads to a nice set of great circles, whose neighbouring circles are along parastiches on the base. The next step is to discretize the fibres into a regular polygon in such a way that the intrafibre distance is close to the interfibre distance, meaning that this discretization depends on the number of points on the base $\mathbb{S}^2$ (chosen here to be a Fibonacci number $F_j$) . Furthermore, a better packing is obtained whenever sites on neighbouring fibres are dephased in such a way that one site on a fibre sits close to the interstices between two sites of the neighbouring fibre. This can be mapped onto an optimization problem directly treated on the base $\mathbb{S}^2$, with one phase factor per base point (and therefore per fibre) and an antiphase preferred configuration between neighbouring sites. Notice that some care should be taken here, due to curvature effects, which already induce some phase difference between distant fibres.  This problem, which is probably frustrated, would deserve a detailed consideration. We did not fully solve it here, and used some approximate phase and discretization values to produce some $P(\mathbb{S}^3,F_j)$ sets, whose coordinates read as follows : 

\begin{eqnarray}
q_1(s,j) &=& \left( \frac{n-s}{n} \right)^{1/2} \exp [ i (\phi_s + \pi  ( s \lambda + 2 j /f)) ] \nonumber \\
q_2(s,j) &=& \left( \frac{s}{n} \right)^{1/2} \exp [ i (\phi_s + \pi (- s \lambda +2 j /f ))]
\label{eq:phylloS3}
\end{eqnarray}

In the above expression, $s$ denotes the fibre, $j$ denotes the site on a fibre (each fibre containing $f$ equally spaced sites, and $\phi_s$ denotes the phase associated with a given fibre $s$. 

Once the $P(\mathbb{S}^3,F_j)$ set is generated, one can display part of it in $\mathbb{R}^3$ by doing a local mapping (either orthogonal or stereographic). One such set is shown in Fig~\ref{fig:phylloS3R3}. Notice that upon using a big Fibonnaci number on the $\mathbb{S}^2$ base, we can reach  very large $P(\mathbb{S}^3,F_j)$ sets, which therefore should be locally very close to the flat space $P(\mathbb{R}^3)$ sets. Indeed, we have studied above a unit sphere $\mathbb{S}^3$ Hopf fibration; but upon rescaling the coordinates with respect to the first neighbour distance between sites, the $\mathbb{S}^3$ radius grows with $F_j$, and the curvature gets smaller and smaller.

\section{Phyllotaxy in $\mathbb{R}^4$}

A quite simple example for a $P(\mathbb{R}^4,s)$ phyllotaxy set is provided by the direct product of two  $P(\mathbb{R}^2)$ sets. Writing coordinates in $\mathbb{R}^4$ as pairs of complex numbers $(q_1,q_2)$, one generate a $P(\mathbb{R}^4,s_1,s_2)$, labelled by the two integers $(s_1,s_2)$ :

\begin{eqnarray}
q_1(s_1+r) &=&  \sqrt{s_1+r} \exp  (2 i \pi (s_1+r) \lambda) \nonumber \\
q_2(s_2+r) &=& -\sqrt{s_2+r} \exp  (2 i \pi (s_2+r) \lambda)
\label{eq:phylloR4}
\end{eqnarray}

Here the quantity $r$ is a shift along the generative spiral (introduced  here to avoid sphere overlap at the origin), and the global minus sign in $q_2$, which could also be avoided, has a particular influence on the structure that will be described below.

As a next step, we may also ask whether interesting $3D$ phyllotactic sets can be derived as $3D$ cuts in 4-dimensional phyllotaxy sets. Notice that this is already true in the crystalline case, where fcc close packed structures can be obtained as cuts perpendicular to the $(1,1,1,1)$ main diagonal in a 4-dimensional hypercubic lattice.

\begin{figure}[h]
\includegraphics[width=1.0\textwidth]{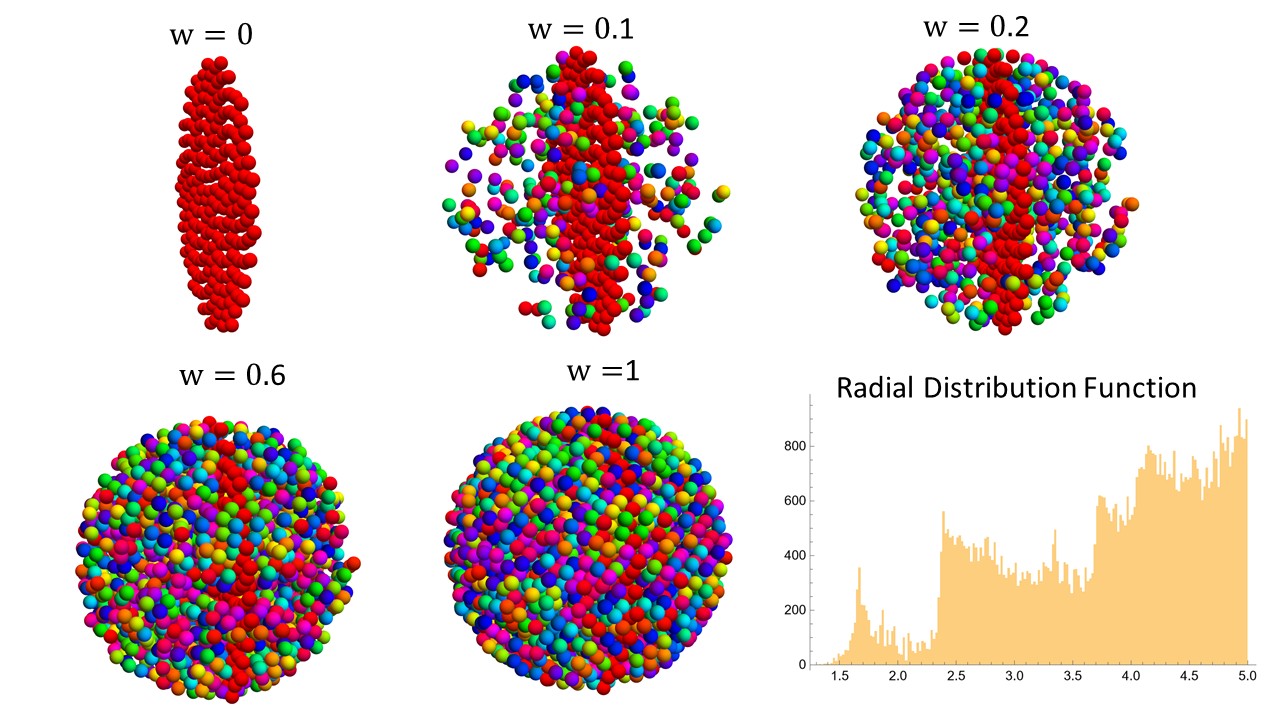}\hfill
\caption{$3D$ phyllotaxy set generated with a codimension one ``Cut and Project" procedure from a a product of $2D$ phyllotaxy sets in $\mathbb{R}^4$. In that case the window $W$ is just a segment, centred at the origin, and of total width $w$ in the $(1,1,1,1)$ direction. This shows the clusters (whose sites are kept inside a ball of radius $R=20$ in $\mathbb{R}^3$) that arise from varying the width "w". Top left , the $w=0$ case (therefore  a simple cut), leading to $2D$ phyllotaxy set in the $(Y,Z)$ plane. The choice of a maximal width is discussed in the text. The colour code (from red to blue)  is related to the distance from the origin in the selection window.}
\label{figphylloCP}
\end{figure}
%

\subsection{Cut and project construction of a $\mathbb{R}^3$ phyllotaxy set}

 But the most important occurrence of higher dimensional ($D+d$) lattices concerns the generation of $D$-dimensional quasiperiodic tilings with the ``Cut and project" (CP) approach (see Ref.\cite{Duneaukatz1985,kaluginkitaevlevitov1985,elser1986}. In that case points are selected in the higher $D+d$ dimensional space according to whether they project inside a given selection ``window" $W$ in the orthogonal $d$-dimensional subspace ($d$ is called the ``codimension"). The precise shape of the selection window is dictated by symmetry arguments and the expected tiling element. 

Here we apply the CP approach in a codimension-one framework (in that case $W$ is simply an interval of width $w$), by selecting out, for the sake of simplicity, the $(1,1,1,1)$ direction. 
The situation is different from the qusicrystal case since we do not start from a crystal in $D+d$ dimensions, and a precise analysis remains to be done . Our aim is to present here an illustrative example. 
We operate a basis change in $\mathbb{R}^4$, leading to coordinates $(X,Y,Z,T)$ :

\begin{eqnarray}
X(s_1,s_2)&=& (\rm{Re}(q_1) - \rm{Im}(q_1) + \rm{Re}(q_2) - \rm{Im}(q_2))/2 \nonumber \\
Y(s_1,s_2) &=& (\rm{Re}(q_1) + \rm{Im}(q_1) - \rm{Re}(q_2) - \rm{Im}(q_2))/2 \nonumber \\
Z(s_1,s_2) &=& (\rm{Re}(q_1) - \rm{Im}(q_1) -\rm{Re}(q_2)  +\rm{Im}(q_2))/2 \nonumber \\
T(s_1,s_2) &=& (\rm{Re}(q_1) +\rm{Im}(q_1) + \rm{Re}(q_2) +\rm{Im}(q_2))/2
\label{eq:rotationR4}
\end{eqnarray}

Keeping the sites such that $|T|<w/2$ leads to structures with coordinates $(X,Y,Z)$ in $\mathbb{R}^3$, depending on the two positive integers $(s_1,s_2)$. Fig~\ref{figphylloCP} shows how the window width influences the structure generation. The case $w=0$ (a simple $3D$ cut in 4D) leads to a simple $2D$ phyllotaxis (scaled by a factor $\sqrt{2}$. This is clearly seen from Eq.~\ref{eq:phylloR4} and Eq.~\ref{eq:rotationR4},  where $X=0$ in the diagonal $(s_1,s_1)$ and $Y(s_1,s_1)$ and $Z(s_1,s_1)$ generates a $2D$ phyllotaxy set. The irrationality of $\lambda$ prevents other $(s_1,s_2)$ pairs to lead $T=0$.
Increasing the acceptance width $w$, more and more points are selected. In the standard CP algorithm, one would tailor the acceptance domain such as to have certain type of accepted tiles, and prevent  having sites too close to existing ones. This usually leads to an acceptance window of precise shape and size. Here, in the absence of  the tiles requirement, we simply use at this step a global threshold (fixing $w$) on the intersite distance for the selected sites. Also shown is the radial distribution function (all distances between all sites) at the last step, with $w=1$.

\subsection{Recovering the $\mathbb{S}^3$ phyllotaxy sets}

The label $(s_1,s_2)$ being positive integers, we can represent $(P(\mathbb{R}^4, s_1,s_2)$ points on the positive quadrant square grid with axes $s_1$ and $s_2$. 
 Interestingly, points in the antidiagonal defined by a $s_1+s_2=R^2$, with $R^2$ any positive integer, belong to concentric $\mathbb{S}^3$ spherical shells of radius $R$ (we have set the shift $r=0$). 

It is therefore tempting to see if we can connect the previously defined $P(\mathbb{S}^3, F_j)$ point sets with shelling in the $P(\mathbb{R}^4, s_1,s_2)$ sets.
This is indeed possible, but with a slightly tricky modification of the 4D construction, which we now briefly describe.

We define rotated $j$-dependant  $P(\mathbb{R}^4,s,j)$ phyllotaxy sets as

\begin{eqnarray}
q_1(s_1) &=&  \sqrt{s_1} \exp (2 i \pi j/f) \exp  (- i \pi s_1 \lambda) \nonumber \\
q_2(s_2) &=&\sqrt{s_2} \exp (2 i \pi j/f) \exp  ( -i \pi s_2 \lambda)
\label{eq:phylloR4S3}
\end{eqnarray}

Note that each of the $s$-dependent phase if half the phase used in Eq.~\ref{eq:phylloR4}. Setting $s_1=R^2-s_2$ leads to a $\mathbb{S}^3$ shelling of the $P(\mathbb{R}^4,s,j)$. If one scales down  the coordinates by a factor $R$, therefore getting a $\mathbb{S}^3$ shell of unit radius, one recovers (up to an unimportant phase) the $\mathbb{S}^3$ above-described phyllotaxy set.

\section{Conclusion}

In this paper, we have proposed several plausible generalizations in $3D$ of the well-known $2D$ phyllotaxy. In the $2D$ case, the key elements are as follows:
\begin{itemize}
\item
A discrete deterministic dynamical system (with parameter $s$) which generates the phyllotaxy set
\item
A radial growth proportional to $\sqrt{s}$, which is a prerequisite for homogeneity
\item
An irrational rotation (based on the golden mean), which allows for global anisotropy
\item
Repetitive (but not  periodic) neighbouring patterns for the site along parastiches spirals, characterized by Fibonacci numbers
\end{itemize}

A direct generalization would have been to implement an automatic discrete set on  $3D$ generative spiral. But, in most cases, we did not find a simple algorithm displaying in $3D$ the generic $2D$ phyllotaxy features. Our examples require generically two parameters ($s_1,s_2$) instead of one. The radial parameter is taken to grow as $s^{1/3}$ to fulfill the homogeneity prerequisite. A notable exception is provided by the deterministic radial packing described in Sec.~\ref{deterministic}, where clusters are described by one varying parameter $s$, at the price of selecting an adapted (modulo triggered) site reordering. 

A more detailed analysis of the generated sets remains to be done, to analyse for instance the degree of homogeneity in the generated clusters, their local stability under a relaxation procedure with a suitable pair potential, and a fine analysis of the local order (via a Voronoi construction) in order to clarify to possible generalization of the parastiche concept in $3D$.

\begin{appendix}

\section{Improved shell-like patterns}

In Section \ref{deterministic}, we obtained a quite interesting shell-like pattern by choosing a reordering that orders sites along the longitudes in the template. We would like to show here that a slightly modified generation rule leads to an improved  (molluscan or sea) shell geometry shape. In addition, a one-parameter family of structures is generated, which leads to a sort of unfolding of the shell, while keeping a $2D$ substructure.
New coordinates read as follows:

\begin{eqnarray}
q(s) &=& s^{1/3} \sin (\theta +\alpha) \,  \exp  (2 i \pi \phi) \nonumber \\
z(s)&=& s^{1/3} \cos (\theta +\alpha) \nonumber \\
{\rm with} \quad \theta &=& \pi \, {\rm frac} \,(s \lambda) \; {\rm and} \quad
 \phi= 2\, n\, \theta \lambda^2   
\label{eq:phylloS2R3b}
\end{eqnarray}

\begin{figure}[h]
\begin{center}
\includegraphics[width=1.0\textwidth]{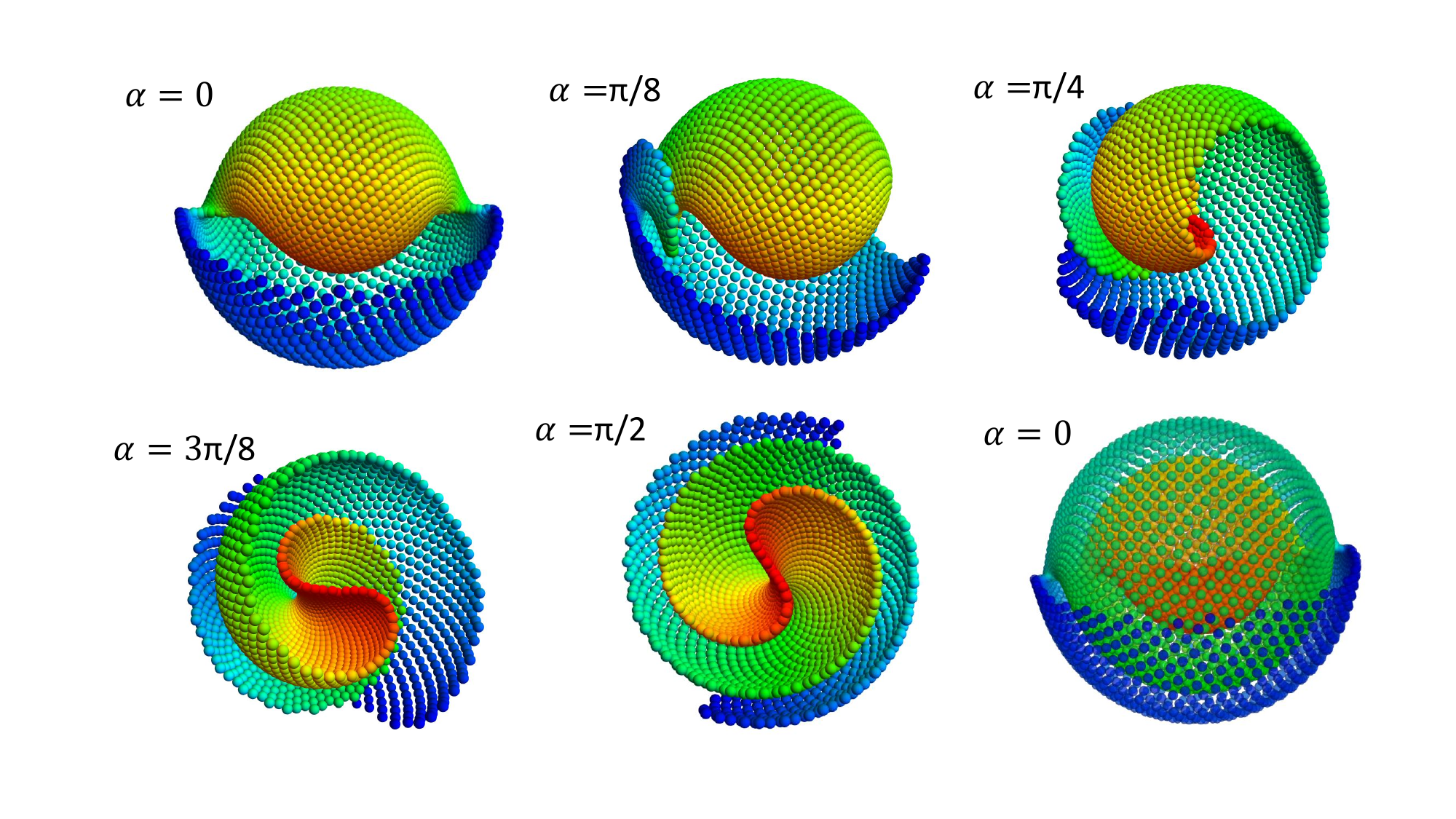}
\end{center}
\caption{Structures generated from Eq.~\ref{eq:phylloS2R3b}, with five values of $\alpha$ from $\alpha=0$ to $\pi/2$.  The colour code goes from red to blue while increasing $s$.  Also shown is a double shell structure in the case $\alpha=0$ where the range of $s$ is doubled.}
\label{fig:phylloS2R3b}
\end{figure}

Figure~\ref{fig:phylloS2R3b} displays five steps, from the shell ($\alpha=0$) to the unfolded one ($\alpha=\pi/2$. In addition the $\alpha=0$ case is also shown by doubling the number of sites, leading to a double shell structure.
%

\end{appendix}

\vspace{1cm}

{\bf{Corresponding author}} : remy.mosseri@cnrs.fr

\vspace{2cm}
\begin{center}
{\bf{Declarations }} 

\end{center}


{\bf{Funding }} : The authors declare that no funds, grants or other support were received during the preparation of this manuscript.

{\bf{Conflict of Interest }} :  The authors have no relevant financial or non-financial interests to disclose.

{\bf{Data Availability }} : The datasets generated during and/or analysed during the current study are available from the corresponding author on reasonable request.

\end{document}